# Quantum Transport in Metal-Porphyrins. An *ab initio* Green's function approach to nanosensors design

Ilya Yanov[a], Juan José Palacios[b] and Glake Hill[a*]

[a] Computational Center for Molecular Structure and Interactions (CCMSI), Department of Chemistry, Jackson State University, MS 39217, United States
[b] Departamento de Física Aplicada, Universidad de Alicante, Alicante 03690, Spain
* Corresponding author: glakeh@ccmsi.us

**Abstract**
An *ab initio* Green's function study of the electron transport properties of the selected metal-porphyrin complexes has been performed. Transmission spectra and current-voltage dependence have been calculated for the porphyrin molecule located between gold electrodes in the presence of interaction with metal atoms, which are most common in biochemistry (Fe(II), Fe(III), Mn(II), and Zn(II)).

It was shown that the estimated Fermi level almost coincides with the LUMO level of Fe(II)-porphyrin and Fe(III)-porphyrin, resulting in significant conductance at small voltage biases. Conductance of Mn(II)-porphyrin and Zn(II)-porphyrin are much lower, and decrease from Mn to Zn.

It was confirmed that performing spin-unrestricted calculations is essential to account splitting of the original molecular orbitals levels.

Preformed calculations demonstrate the principal possibility of experimental realization of porphyrin based nanosensors.

**Introduction**
Recent developments in nanotechnology open the possibility for production of nanoscale sensors that provide instant and inexpensive way to monitor environment conditions and to diagnose chemical and biological hazards. One of the widely proposed molecules for design of nanosensors is the porphyrin (e. g. US Pat. No. 5,981,202; 6,402,0376; 6,488,891; 6,495,102). Porphyrins are nitrogen-containing compounds derived from the parent tetrapyrroleporphin molecule. Metalloporphyrins play an important role in biology. Chlorophyll is a Mg-porphyrin and the Fe(II) porphyrin complex is the part of hemoglobins and myoglobins, which are responsible for oxygen transport and storage in living tissues. An important feature for practical applications is that the porphyrin rings are very stable to influence of concentrated acids and they itself can act both as an acid and a base [1].

Most of the proposed applications relied on spectroscopic methods for detecting the interaction between the target molecule and a sensor. The absorption spectrum of porphyrin shows Soret intensive absorption band at around 400 nm in near-ultraviolet spectra, followed by several weaker absorptions (Q Bands) at higher wavelengths (450 to 700 nm) in visible spectra. Variations of the peripheral substituents of the porphyrin ring cause changes to the intensity and wavelength of these absorptions. Protonation of two of the inner nitrogen atoms or insertion of a metal into the porphyrin cavity also changes the visible absorption spectrum [2]. The wavelength maxima of the absorbance spectrum and the extinction coefficient (absorptivity) of metallo-porphyrins are affected by neighboring molecules. Factors causing an increase in electronic density of pi-electron orbitals at the periphery of the porphyrin tend to cause red shifts of the absorbance bands. Red shifts are found to arise as a function of the electron affinity of side chain substituents [3-5].

These features can facilitate the development of nanoscale sensor units, but fully operational device will require spectroscopic analytical equipment. A perspective new way for the design of nanosensors is to incorporate porphyrin molecules in electric circuits and obtain information amperometrically, by measuring electric current that results from interaction between porphyrin and the agent molecule. The aim of our work is the *ab initio* non-equilibrium Green's function study of the electron transport properties of the metal-porphyrin complexes.

**Methods and calculations**
The main feature of electronic devices is that they are open systems with respect to electron flow. A theoretical consideration of such devices should be done in terms of statistically mixed states and quantum kinetic theory [6]. The only completely adequate form of this theory is non-equilibrium Green's function formulation of many-body theory. In this paper we use the ALACANT code [7], which is based on the non-equilibrium Green's function formalism and works as an interface to GAUSSIAN program sets [8]. It allows to calculate the electronic structure and conductance on the basis of density functional theory. ALACANT divides the space into left (L) and right (R) semi-infinite electrode regions and a contact (C) region which includes the description of electrode surfaces as well as the atom-scale conductor. It employs a Bethe lattice model, appropriately parametrized to mimic the bulk density of states of non-crystalline metallic electrodes, to build the self-energy matrices that are coupled to the Hamiltonian matrix of the C region. This mean-field-like, one-electron approach is not able to describe many-body effects which may appear in some cases during the transport process. Inelastic scattering by phonons will not be considered either in our study.

**Results and discussion**

The scheme of the electric circuit where a porphyrin molecule is chemically bound to the Au model contacts is shown in Figure 1. Initial geometries were optimized at the B3LYP/STO-3G and B3LYP/6-31G(d) level of theory using the GAUSSIAN98 [8] program set. Christiansen core pseudotentials described in Refs. [9-11] were used for Au atoms.

$G(E_f) = (2e^2/h)T(E_f)$, where $T(E_f)$ denotes the total transmission of all the conductance channels at the Fermi energy. Conductance of Mn(II)-porphyrin and Zn(II)-porphyrin are much lower (Figure 2c,d), and decrease from Mn to Zn.

It should also be noted that performing spin-unrestricted calculations is essential to obtain the correct splitting of the original molecular levels as the orbitals get filled or emptied by the electrodes in the self-consistent alignment of the molecular levels and the Fermi energy of the metal [13]. This is of particular relevance when the molecular orbitals are weakly coupled to the electrodes, as is the case here, and charge localization is strong (note that the use of hybrid functionals here like B3LYP is crucial to avoid self-interaction effects [13]).

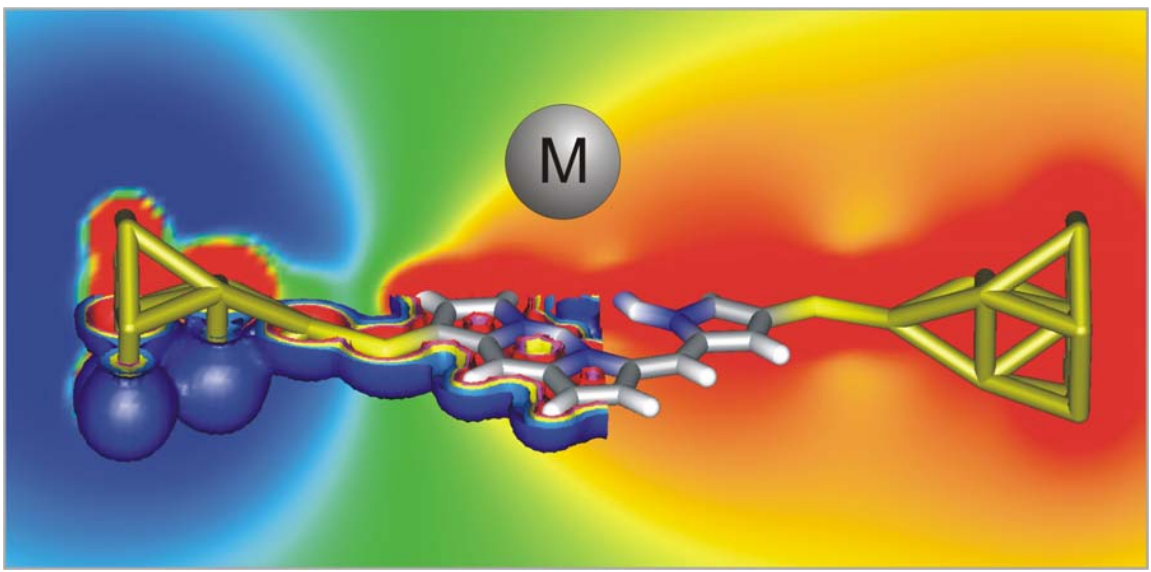

Figure 1. Scheme of the metal-porphyrin complex attached to the Au electrodes.

It is known that the sulfur atoms create strong covalent bonds with the Au (111) surface in the tetrahedral position, where the distance to the nearest gold atoms is 2.53 Å. [12]. Therefore the terminal hydrogen atoms of porphyrin molecule were substituted by the sulfur atoms to facilitate bonding of the porphyrin molecule to the gold surface.

The results of the calculations of transmission spectra for the considered systems are presented in Figure 2. It is shown (Figure 2a,b) that because the self-consistent Fermi level of the system almost coincides with the LUMO level of molecules, Fe(II)-porphyrins and Fe(III)-porphyrins exhibit significant conductance at small voltage. According to Landauer's formalism,

**Summary**

In this paper we have discussed the electron transport properties and the possibility of utilizing the porphyrin molecule as a molecular electronic device which can be applied as part of a biosensor, molecular switch, or resonant tunneling diode. Calculations were performed using DFT theory and non-equilibrium Green's function formalism with the help of the ALACANT code.

It was shown that the estimated Fermi level almost coincides with the LUMO level of Fe(II)-porphyrin and Fe(III)-porphyrin, resulting in significant conductance at small voltage biases. Conductance of Mn(II)-porphyrin and Zn(II)-porphyrin are much lower, and decrease from Mn to Zn. Such strong dependence of conductivity of the system in the presence of interaction with different metal atoms, demonstrate the principal possibility of experimental realization of porphyrin based nanosensors.

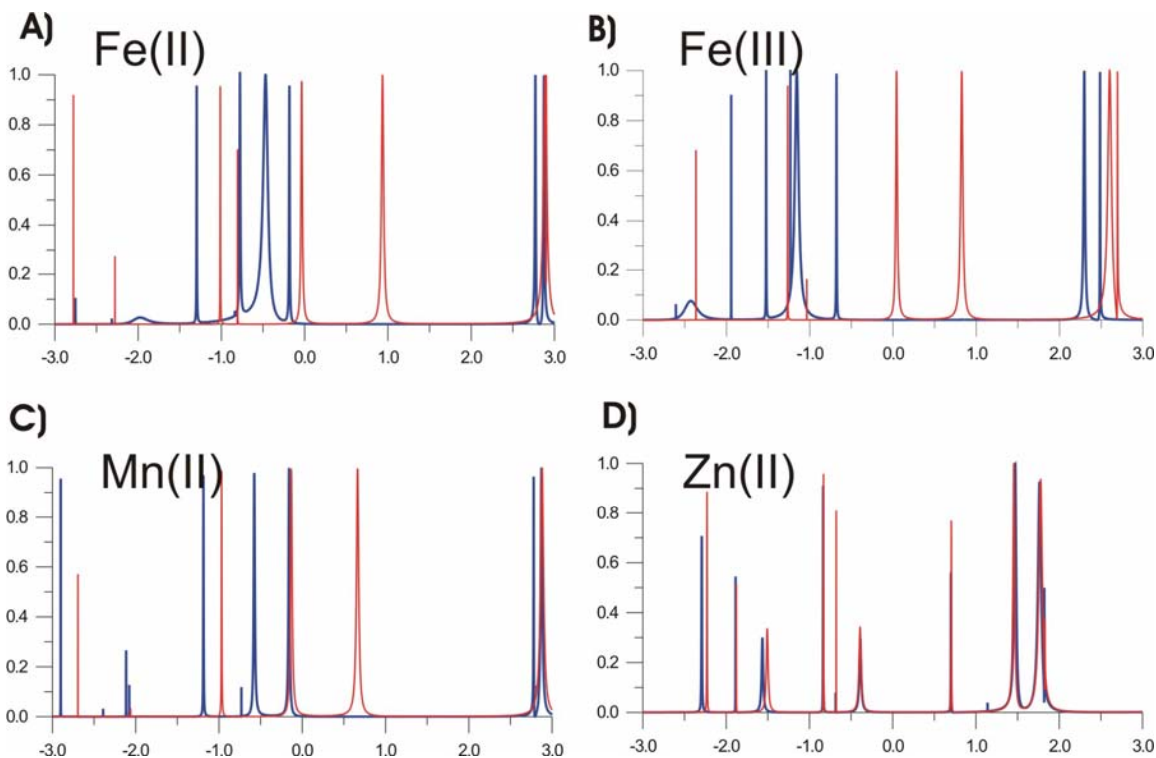

Figure 2. Transmission spectra of metal-porphyrins between gold electrodes. Blue color corresponds to alpha electrons; red color indicates beta electrons.

The importance of spin-unrestricted calculations has been shown. Spin-restricted models significantly overestimate conductance.

**Acknowledgements**
This work was supported by the Department of Defense through the U. S. Army Engineer Research and Development Center (Vicksburg, MS), Contracts #W912HZ-04-2-0002 and #W912HZ-05-C-0051, and NSF-PREM Grant # DMR-0611539. J.J.P. acknowledges financial support from MEC of Spain under Grant No. MAT2005-07369 and from the University of Alicante.

**References**
1. Porphyrins and Metalloporphyrins. K. M. Smith, Ed., Elsevier, Amsterdam, 1-910 (1975).
2. Goldoni, A. Porphyrins: fascinating molecules with biological significance; Elettra highlights; Trieste, 2001-2002; 64-65.
3. Igarashi, S.; Yotsuyanagi, T. Analytica Chimica Acta 1993, 281, 347-351.
4. Mauzerall, D. Biochem. 1965, 4, 1801-1810.
5. Karasevich, I. E.; Anisomova, B. L.; Rubailo, V. L.; Shilov, A. E. Kinetics and Catalysis, 1993, 34, 651-657.
6. Frensley, W. R. Quantum Transport; Academic: San Diego, 1994; Chapter 2.
7. Palacios, J. J.; Pérez-Jimenez, A. J.; Louis, E.; Vergés, J. A. Phys Rev B 2001, 64, 115411; Palacios, J. J.; Pérez-Jimenez, A. J.; Louis, SanFabián, E.; E.; Vergés, J. A. Phys Rev B 2002, 66, 035322; for more information on the code, visit the webpage: www.guirisystems.com/alacant
8. GAUSSIAN98, Revision A.7, Gaussian, Inc., Pittsburgh PA (1998).
9. Pacios, L. F.; Christiansen, P. A. J. Chem. Phys. 82, 2664 (1985).
10. Hurley, M. M.; Pacios, L. F.; P. A. Christiansen, R. B. Ross, and W. C. Ermler. J. Chem. Phys 84, 6840 (1986).
11. R. B. Ross, J. M. Powers, T. Atashroo, W. C. Ermler, L. A. LaJohn, and P. A. Christiansen. J. Chem. Phys. 93, 6654 (1990).
12. A. I. Yanson, G. R. Bollinger, H. E. van den Brom, N. Agrait, and J. M. Van Ruitenbeek. Nature, 395, 783 (1998).
13. J. J. Palacios, Phys. Rev. B 72, 125424 (2005).